\documentclass[iop]{emulateapj}
\usepackage{apjfonts, natbib}
\usepackage{amsmath}
\usepackage{graphicx}
\usepackage{bm}

\def\be{\begin{equation}}
\def\ee{\end{equation}}
\def\ba{\begin{eqnarray}}
\def\ea{\end{eqnarray}}
\def\nn{\nonumber}

\def\mA{\mathbf{A}}
\def\vb{\mathbf{b}}
\def\vn{\mathbf{n}}
\def\vtheta{\bm{\theta}}
\def\vgamma{\bm{\gamma}}
\def\vepsilon{\bm{\epsilon}}
\def\lsim{\mathrel{\rlap{\lower4pt\hbox{\hskip1pt$\sim$}}
    \raise1pt\hbox{$<$}}}               
\def\gsim{\mathrel{\rlap{\lower4pt\hbox{\hskip1pt$\sim$}}
    \raise1pt\hbox{$>$}}}    
\renewcommand\Re{\operatorname{Re}}

\begin{document}

\title{Mapping the dark matter with polarized radio surveys}
\shorttitle{Mapping the dark matter with polarized radio surveys}

\author{Michael L. Brown$^{1,2}$ and Richard A. Battye$^{3}$}
\affil{$^{1}$Astrophysics Group, Cavendish Laboratory, University of
  Cambridge, J J Thomson Avenue, Cambridge CB3 OHE, United Kingdom\\
$^{2}$Kavli Institute for Cosmology, University of
  Cambridge, Madingley Road, Cambridge CB3 OHA, United Kingdom\\
$^{3}$Jodrell Bank Centre for Astrophysics, School of Physics and
  Astronomy, University of Manchester, Oxford Road, Manchester, M13
  9PL, United Kingdom}
\shortauthors{M. L. Brown and R. A. Battye}

\begin{abstract}
In a recent paper \citep{brown11}, we proposed the use of integrated
polarization measurements of background galaxies in radio weak
gravitational lensing surveys and investigated the potential impact on
the statistical measurement of cosmic shear. Here we extend this idea
to reconstruct maps of the projected dark matter distribution, or
lensing convergence field. The addition of polarization can, in
principle, greatly reduce shape noise due to the intrinsic dispersion
in galaxy ellipticities. We show that maps reconstructed using this
technique in the radio band can be competitive with those derived
using standard lensing techniques which make use of many more
galaxies. In addition, since the reconstruction noise is uncorrelated
between these standard techniques and the polarization technique,
their comparison can serve as a powerful check for systematics and
their combination can reduce noise further. We examine the convergence
reconstruction which could be achieved with two forthcoming
facilities: (i) a deep survey, covering $1.75$ deg$^2$ using the
e-MERLIN instrument currently being commissioned in the UK and (ii) the
high resolution, deep wide field surveys which will eventually be
conducted with the Square Kilometre Array (SKA).
\end{abstract}


\section{Introduction}
\label{sec:intro}
Weak gravitational lensing has become one of the most powerful
techniques for investigating the distribution of dark matter on
cosmological scales (e.g. \citealt{massey07, heymans08,
schrabback10}). The majority of weak lensing studies to date have been
conducted in the optical bands since large numbers of galaxies are
required to reduce the shape noise associated with the intrinsic
ellipticities of galaxies. Observationally, the pace of progress in
the optical is set to continue with forthcoming wide-field surveys
using a number of purpose-built instruments, 
e.g.~DES\footnote{\url{http://www.darkenergysurvey.org}},
KIDS\footnote{\url{http://www.strw.leidenuniv.nl/~kuijken/KIDS/}}, 
Euclid\footnote{\url{http://sci.esa.int/science-e/www/area/index.cfm?fareaid=102}},
WFIRST\footnote{\url{http://wfirst.gsfc.nasa.gov/}} and 
Pan-STARRS\footnote{\url{http://pan-starrs.ifa.hawaii.edu}}.

As the statistical precision of optical surveys improves, greater
control over potential systematics is required. Major systematics of
concern include instrumental effects such as anisotropies in the
point-spread function (PSF) of the telescope and astrophysical
systematics such as intrinsic galaxy alignments
(e.g. \citealt{crittenden01, catelan01, hirata04}). 

In addition to a new generation of wide-field optical surveys, a suite
of powerful next-generation radio instruments are due to come online
in the near future. Radio instruments offer a number of potential
advantages over optical surveys for lensing studies. In particular,
radio interferometers do not suffer from complicated PSF effects,
their synthesized beams being precisely and solely determined by the
telescope array configuration and direction of observation.

A further unique advantage offered by measuring lensing in the radio
band is the polarization information which is usually measured in
addition to the total intensity in radio surveys. Previous authors
have exploited the fact that the polarization position angle is
unaffected by lensing in order to measure gravitational lensing of
distant quasars \citep{kronberg91, kronberg96, burns04}.  In a recent
paper \citep{brown11}, we showed how one could extend this idea to
measure cosmic shear. The technique relies on there existing a
reasonably tight relationship between the orientation of the
integrated polarized emission and the intrinsic morphological
orientation of the galaxy. The existence of this relationship needs to
be established for the high-redshift star-forming galaxies which are
expected to dominate the radio sky at the $\mu$Jy flux sensitivities
achievable with forthcoming instruments. However, such a relationship
certainly exists in the local Universe \citep{stil09} and it is
reasonable to assume that it persists to higher redshift.

A key difference between the polarization technique and standard
techniques for measuring lensing is that the former \emph{does not}
assume that the ensemble average of the intrinsic shapes of galaxies
vanishes. It is thus, in principle, able to cleanly discriminate
between a lensing signal and a possible contaminating signal due to
intrinsic galaxy alignments \citep{brown11}. 

In this letter, we investigate the direct reconstruction of the
projected dark matter distribution (or lensing convergence field)
using the radio polarization technique. Our simulations are based on
the ray-tracing simulations of \cite{white05}. We begin in
Section~\ref{sec:sim} by summarizing the salient features of these
simulations and we describe the parameters we have adopted to mimic
the e-MERLIN\footnote{\url{http://www.e-merlin.ac.uk}} (currently
being commissioned with full operation planned for the second half of
2011) and SKA\footnote{\url{http://www.skatelescope.org}} instruments. In
Section~\ref{sec:method}, we describe how to include polarization
information in a direct-inversion mass reconstruction algorithm and we
present our simulation results. We finish in Section~\ref{sec:discuss}
with a discussion.

\section{Simulations}
\label{sec:sim}
To demonstrate the mass reconstruction technique using polarization
information, we have made use of the simulated lensing convergence and
shear maps of \cite{white05}\footnote{Available at
\url{http://mwhite.berkeley.edu/Lensing/Thousand/}}. These simulations
comprise $\approx 1000$ square degrees of simulated sky from high
resolution numerical simulations based on a $\Lambda$CDM cosmology
with parameters, $\Omega_m = 0.28, \Omega_bh^2 = 0.024$, $h=0.7$,
$\sigma_8=0.9$ and $n_s = 1$. The simulations consist of 112 $3 \times
3$ deg$^2$ fields which are approximately independent.

The assumed redshift distribution is of the form, $n(z) \propto z^2
\exp(-z/z_0)^\beta$ with $z_0 = 1.0$ and $\beta = 1.5$. The median
redshift of this distribution is $z_m = 1.41$. Making use of the
semi-empirical SKA Design Studies (SKADS) simulation of
\cite{wilman08}\footnote{See \url{http://s-cubed.physics.ox.ac.uk/}},
we have checked that this $n(z)$ agrees reasonably well with the
redshift distribution of star-forming galaxies expected in future deep
radio surveys. In particular, we measure $z_m = 1.02$ in the SKADS
simulation for a $10\sigma$ detection threshold of $40$ $\mu$Jy, a
sensitivity which could be reached over 1.75 deg$^2$ using e-MERLIN.
Large-scale surveys with the SKA might go $\sim10$ times deeper
(e.g.~\citealt{blake07}). At this depth, we measure a median redshift of $z_m =
1.48$ from the SKADS simulation. Although the predicted median
redshift of the proposed e-MERLIN survey is somewhat smaller than that
of White's simulations, this discrepancy is sub-dominant when compared
to the current very large uncertainties in predicting $n(z)$ for
$\mu$Jy-sensitivity radio surveys.

A further unknown for forthcoming radio surveys is the number density
of galaxies that will be achieved. Source counts at 1.4~GHz are
presented in \cite{biggs06} for deep fields over relatively small
areas of sky. Extrapolating from this study suggests a source density
of $\sim1$ arcmin$^{-2}$ for a detection threshold of $40$
$\mu$Jy. This source density is also consistent with that found in
deep VLA + MERLIN observations of the Hubble Deep Field
\citep{muxlow05}. The source counts from a more recent study by
\cite{owen08} are significantly higher and suggest a source density of
$2$ arcmin$^{-2}$ could be achieved. For the e-MERLIN simulations, we
take the average of these two estimates and adopt $\bar{n} = 1.5$
arcmin$^{-2}$ for the number density of galaxies detected in total
intensity. For the SKA simulations, we assume $\bar{n} = 15$
arcmin$^{-2}$.

The polarization properties of the high-redshift star-forming galaxies
which will dominate these surveys are currently unknown. We can only
extrapolate from the local Universe where the fractional polarization
is typically between $0$ and $20$\% \citep{stil09}. These observations
of local galaxies also suggest that the polarized emission and the
morphological orientation are well aligned to an accuracy of between
$0$ and $\sim15$ degs. 

For our simulations, we assume that the galaxies are, on average, 10\%
polarized. Noting that a measurement of the orientation of the
polarized emission requires only a third of the sensitivity which
would be required to measure the shape of the emission, we assume
further that we can measure a useful polarization orientation for one
third of the galaxies detected in total intensity. Thus, we assume
$\bar{n}_{\rm pol} = 0.5$ arcmin$^{-2}$ for e-MERLIN and $\bar{n}_{\rm
pol} = 5.0$ arcmin$^{-2}$ for the SKA. We further assume that such
polarization orientation detections are an unbiased tracer of the
intrinsic morphological orientation of the galaxy with a scatter of
$\alpha_{\rm rms} = 7$ degs. (See Section 2 of \citealt{brown11} for
further discussion on the current observational status for
polarization and possible values for these parameters.)

\begin{figure*}
\begin{center}
\resizebox{1.03\textwidth}{!}{\includegraphics{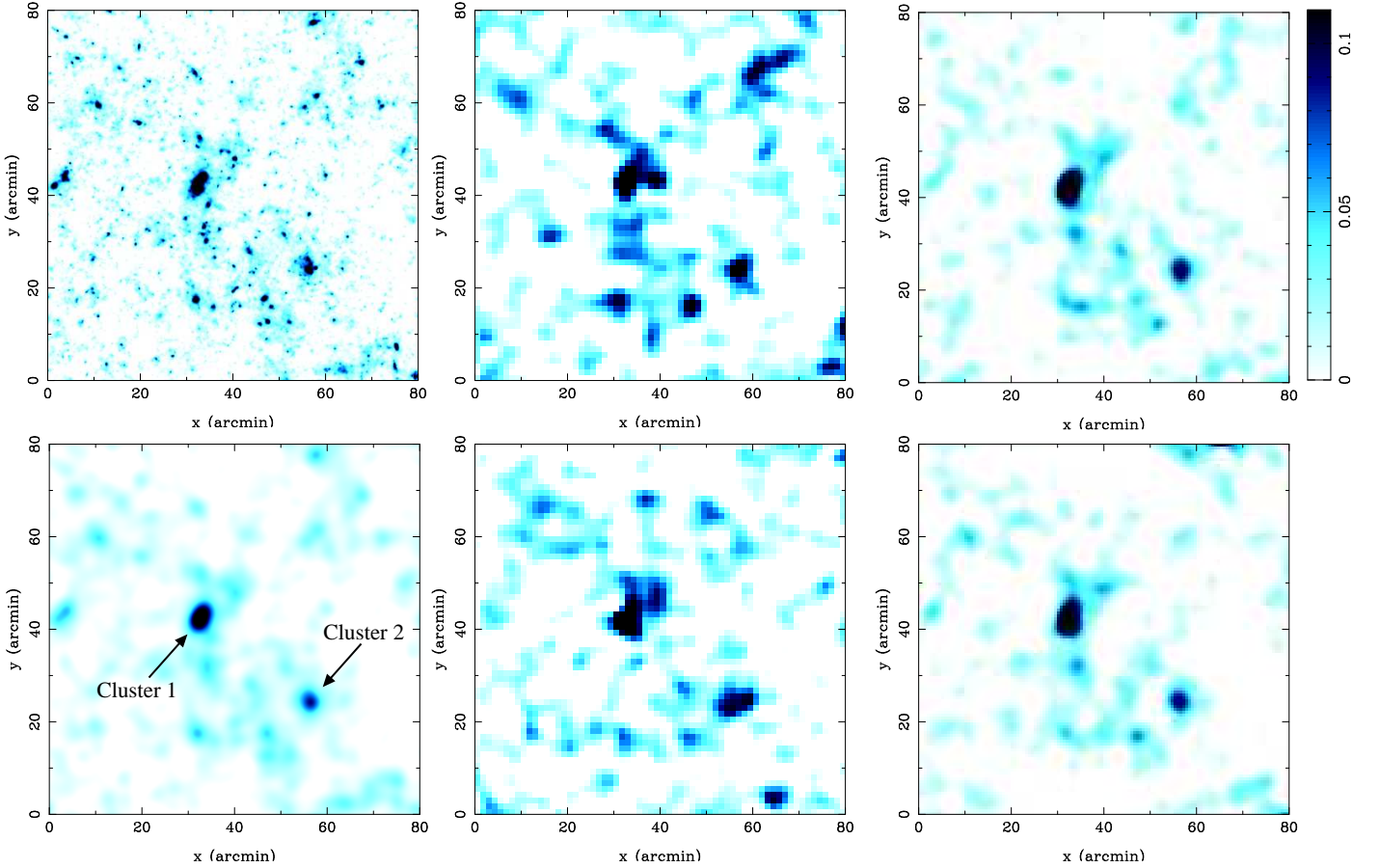}}
\end{center}
\vspace{-0.3cm}
\caption{An example reconstruction of the dark matter distribution in
  a 1.75 deg$^2$ region of the simulations. The input convergence
  field is shown unaltered in the upper left panel and smoothed with a
  2 arcmin Gaussian kernel in the lower left panel. The simulated
  reconstruction using total intensity only for observational
  specifications similar to the proposed e-MERLIN survey (with
  $\bar{n} = 1.5$ arcmin$^{-2}$ and $\epsilon_{\rm rms} = 0.30$) is
  shown in the upper central panel. Assuming one third of these
  galaxies are sufficiently polarized to yield an intrinsic position
  angle tracer that is unbiased with a scatter of $\alpha_{\rm rms} =
  7.0$ degs, the reconstruction obtained using polarization
  information is shown in the lower central panel. The reconstructions
  obtainable with the SKA are shown in the right-hand panels (upper
  panel --- using total intensity only; lower panel --- with
  polarization). For the SKA reconstructions, we have assumed $\bar{n}
  = 15$ arcmin$^{-2}$ and that one third of the galaxies are useable
  in polarization with $\alpha_{\rm rms} = 7.0$ degs. Integrated
  convergence measurements for the two structures labelled ``Cluster
  1'' and ``Cluster 2'' are given in Table~\ref{tab:rms_noise}.}
\label{fig:dm_maps}
\end{figure*}

To mimic the proposed e-MERLIN survey, we extract the central 1.75 deg$^2$
from each $3 \times 3$ deg$^2$ simulation. An example of the
convergence distribution for one field is shown in the top left panel
of Fig.~\ref{fig:dm_maps} where a number of large-scale structures and
filaments are clearly visible. Using the parameters detailed above, we
generate background galaxies randomly distributed across the field of
view. Each galaxy is assigned an intrinsic ellipticity, $\vepsilon =
\epsilon_1 + i\,\epsilon_2$ where $\epsilon_1$ and $\epsilon_2$ are
drawn from zero-mean Gaussian distributions with width $\sigma_\epsilon =
0.30$. For each galaxy, we then add the simulated shear to the
intrinsic ellipticity, interpolating the shear from the numerical
simulations.

For the polarization technique, we additionally simulate the observed
orientation of the polarized emission of each galaxy according to
$\alpha^{\rm obs} = \alpha^{\rm int} + \alpha^{\rm rand}$ where
$\alpha^{\rm int} = (1/2) \tan^{-1}(\epsilon_2/\epsilon_1)$ is the
intrinsic position angle of the galaxy and $\alpha^{\rm rand}$ is the
random noise in the polarized orientation---intrinsic position angle
relationship.

\section{Mass reconstruction using polarization}
\label{sec:method}

To reconstruct the convergence field from the simulated galaxy
catalogues, we use a variant of the Kaiser-Squires direct-inversion
technique (\citealt{kaiser93}). Our implementation is similar to the
real-space algorithm described in \cite{seitz95}. To perform the
reconstruction using the total intensity only (i.e.~using the standard
lensing technique), following \cite{seitz95}, we first estimate the
mean shear field on a regular grid covering the entire field. We
estimate the shear in each grid cell as (the indexes, \{i,j\} denote
the grid cells)
\be
\hat{\vgamma}_{i,j} = \frac{\sum_k w_k \vepsilon^{\rm obs}_k}{\sum_k w_k} 
\ee
where the sum is over all galaxies and we have applied a weight of 
\be 
w_k = \exp\left(-\frac{(\vtheta_{i,j} - \vtheta_k)^2}{(\Delta\theta)^2}\right) 
\label{eqn:weights}
\ee 
to each of the observed ellipticities, $\vepsilon^{\rm obs}_k$. We use
a smoothing scale of $\Delta\theta = 2.0$ arcmin. The convergence
field is then estimated as a convolution,
\be
\hat{\kappa}(\vtheta) = -\frac{1}{2\pi} \sum_{i,j} \Re [ \mathcal{D}(\vtheta
- \theta_{i,j}) \, \vgamma^*(\vtheta_{i,j}) ]
\label{eqn:kappa_est}
\ee
where the kernel, $\mathcal{D}$ is given by 
\be
\mathcal{D} = \frac{\theta_1^2 - \theta_2^2 + 2\,i\, \theta_1
  \theta_2}{\left|\vtheta\right|^4}.
\ee

Including the polarization information in this estimator is
straightforward and simply involves using a weighted version of the
shear estimator described in \cite{brown11} to estimate the shear at each grid
point. Explicitly, the shear in each grid cell is now estimated as
\be
\hat{\vgamma} = \mA^{-1} \vb,
\label{eqn:new_est}
\ee
where the matrix, $\mA$, and the vector, $\vb$, are given by
\ba
\mA &=& \sum_k w_k^{} \, \hat{\vn}_k^{} \, {\hat{\vn}_k}^T, 
\label{eqn:Amatrix_def} \\
\vb &=& \sum_k w_k \, (\vepsilon_k^{\rm obs} \cdot \hat{\vn}_k) \, \hat{\vn}_k.
\label{eqn:bvec_def}
\ea
Here, $\hat{\vn}_k = \left( \sin2\alpha_k^{\rm obs},
-\cos2\alpha_k^{\rm obs} \right)$ is the direction at $45^\circ$
to the estimate of the intrinsic position angle of the galaxy as
provided by the polarization and $w_k$ is a weight. 

Using a weighting scheme in equation~(\ref{eqn:new_est}) is now
crucial -- the small galaxy number density we are dealing with means
that ``shot noise'' is a severe problem for the polarization
estimator.\footnote{We discriminate here between ``shape noise'' which
is the dispersion in the intrinsic shapes of galaxies and ``shot
noise'' which is the noise introduced due to the finite number of
galaxies and hence the sparse sampling of the underlying shear field.}
Had we used an unweighted shear estimator (i.e. simple binning) then a
significant fraction of our grid cells would contain no galaxies for
the polarization estimator and this would present problems for the
convolution operation. By using weights which are non-negligible at
the typical separation of polarized galaxies, we can recover the shear
at each point in the grid, albeit at the expense of a smoothing of the
shear field. In practice, we use the same weighting scheme as for the
standard case (equation~\ref{eqn:weights}) with the same smoothing
scale, $\Delta \theta = 2.0$ arcmin, in all cases. Once the shear
field has been estimated, the convergence, $\kappa(\vtheta)$ is found
as in the standard case (equation~\ref{eqn:kappa_est}).

An example of the mass reconstruction performance for both the
standard estimator and the polarization estimator is shown in
Fig.~\ref{fig:dm_maps}. The simulated e-MERLIN reconstructions have
been performed on a $64 \times 64$ grid of 1.25 arcmin pixels. For the
much higher fidelity SKA simulations, we use a $128 \times 128$ grid
of 0.625 arcmin pixels. We see that the major mass concentrations can
be successfully identified in the simulated e-MERLIN reconstructions
while the SKA maps also trace the filamentary structure of the
numerical simulations extremely well. For the observational parameters
we have adopted, the reconstructions obtained using the polarization
estimator are of similar sensitivity to those obtained using the
standard estimator. In Table~\ref{tab:rms_noise} we list the
reconstructed convergence field integrated in apertures of radius 3
arcmin for the two largest structures in the field, centred on $(x, y)
= (32.6, 42.8)$ arcmin and $(x, y) = (56.5, 24.3)$ arcmin. The errors
are determined by repeating the analysis on empty fields and measuring
the root-mean-square convergence field integrated in identical
apertures.

\begin{deluxetable}{c c c}
\tablecaption{Integrated convergence measurements for the structures
  labelled ``Cluster 1'' and ``Cluster 2'' in Fig.~\ref{fig:dm_maps}.} 
\tablehead{\colhead{Simulation} & \colhead{Standard estimator} & \colhead{Polarization estimator}}
\startdata
e-MERLIN: &                 &                 \\
Cluster 1 & $2.36 \pm 0.19$ & $2.31 \pm 0.18$ \\
Cluster 2 & $1.26 \pm 0.19$ & $1.30 \pm 0.18$ \\
\\  
SKA:       &                 &                 \\           
Cluster 1 & $2.29 \pm 0.08$ & $2.13 \pm 0.07$ \\
Cluster 2 & $0.92 \pm 0.08$ & $0.98 \pm 0.07$
\enddata
\label{tab:rms_noise}
\vspace{0.5cm}
\end{deluxetable}

We have also measured the shear correlation functions from the
simulated maps. The correlation functions are defined as 
\ba
C_1(\vtheta) &=& \langle\gamma_1^r(\vtheta)\gamma_1^r(\vtheta + \Delta\theta)\rangle \nn \\
C_2(\vtheta) &=& \langle\gamma_2^r(\vtheta)\gamma_2^r(\vtheta + \Delta\theta)\rangle \nn \\
C_3(\vtheta) &=& \langle\gamma_1^r(\vtheta)\gamma_2^r(\vtheta + \Delta\theta)\rangle 
\ea
where the angled brackets denote an average over all map-pixel pairs
separated by $\Delta\theta$ and $\gamma^r_1$ and $\gamma^r_2$ are the
estimated shear components rotated to a coordinate system aligned with
the vector joining the two map-pixels.  The cross-correlation,
$C_3(\vtheta)$ is expected to vanish in the absence of
parity-violating effects.

We have measured the total correlation function, $C(\vtheta) =
C_1(\vtheta) + C_2(\vtheta)$ using both the standard and polarization
estimator and have also combined the results from the two
techniques. Note that the combination of the two estimators is
powerful since the dominant source of measurement noise is different
in the two cases --- for the standard estimator, the shape noise
dominates while for the polarization estimator the noise associated
with $\alpha_{\rm rms}$ dominates meaning that there is useful extra
information in the cross-correlation of the shear fields estimated
using the two techniques. The cross-correlation can also be useful in
mitigating systematics which would likely be different
for the two techniques.

\begin{figure}
\vspace{0.5cm}
\begin{center}
\resizebox{0.95\columnwidth}{!}{
  \rotatebox{-90}{\includegraphics{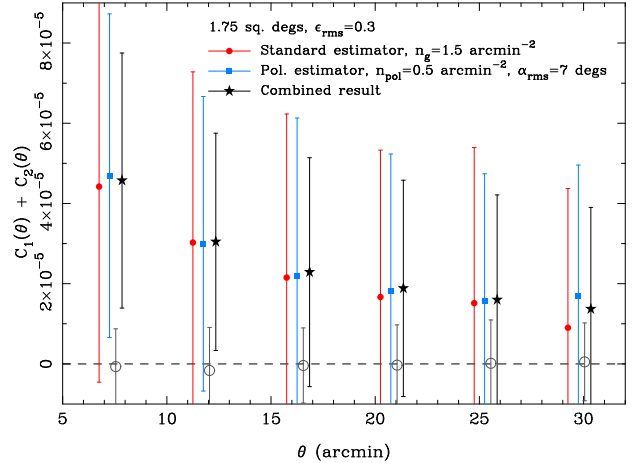}}}
\end{center}
\caption{The total shear correlation function measured from the
  e-MERLIN simulations. Three sets of points are plotted. From left to
  right they have been obtained using the standard shear estimator,
  using the polarization estimator and their combination. The points
  plotted as open circles show the cross-correlation, $C_3(\vtheta)$, which is
  consistent with zero. Comparison of the errors on this latter
  measurement with those on the measurement of the signal gives an
  indication of the relative sizes of the measurement noise and sample
  variance contributions. The errors plotted are those appropriate for
  a single realization. Measurements on different scales are heavily
  correlated and the significance of the detection of cosmic shear is
  in fact only $1.4\sigma$.}
\label{fig:ctheta}
\end{figure}

The results, averaged over simulations, are plotted in
Fig.~\ref{fig:ctheta}. Naively, it looks as though a detection of
cosmic shear is possible with the proposed 1.75 deg$^2$ e-MERLIN
survey. However, the measurements on different scales are strongly
correlated and the formal detection significance is only
$1.4\sigma$. Note that the dominant contribution to the errors is
sample variance. Instead of doing cosmology, if we ask the question
how well can we measure the lensing signal in a typical 1.75 deg$^2$
field, then the significance is much larger ($\sim 6.6\sigma$).

\section{Discussion}
\label{sec:discuss}

In this letter, we have extended the techniques of \cite{brown11} to
the lensing reconstruction of dark matter maps from polarized radio
surveys. For the parameters which we have adopted, the technique looks
extremely promising and could make future radio surveys as powerful as
optical lensing surveys containing many more galaxies.

There is, of course, a great deal of uncertainty with respect to the
values one should adopt for the polarization properties of
star-forming galaxies at $\mu$Jy flux densities. In particular, the
typical fractional polarization and the degree to which the
orientation of the polarized emission is aligned with the intrinsic
morphological orientation are currently unknown. We note also however
that there is a trade-off in these two parameters. For example, one
can imagine including a larger number of galaxies whose polarized
emission and intrinsic orientation are poorly aligned (large
$\bar{n}_{\rm pol}$ and large $\alpha_{\rm rms}$) or one could be very
selective about which galaxies one includes, choosing high fractional
polarization objects whose polarization orientation closely traces the
major axis of the galaxy (low $\bar{n}_{\rm pol}$ and low $\alpha_{\rm
  rms}$).

\begin{figure}
\begin{center}
\resizebox{\columnwidth}{!}{
  \rotatebox{-90}{\includegraphics{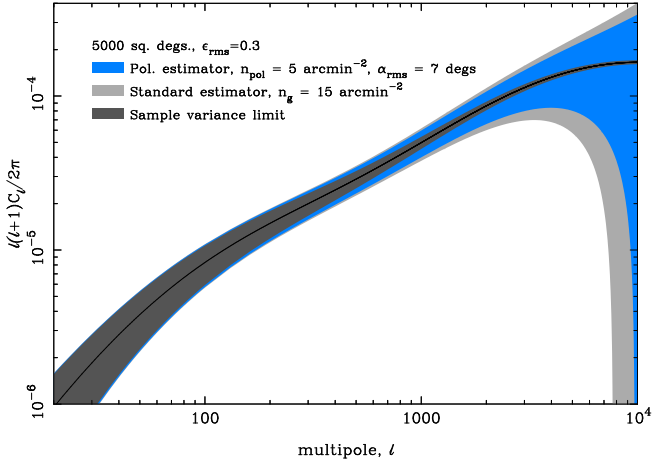}}}
\end{center}
\vspace{-0.3cm}
\caption{Predicted per-multipole uncertainties on the weak lensing
  power spectrum for a 5000 deg$^2$ survey with $\bar{n} = 15$
  arcmin$^{-2}$ and $\vepsilon_{\rm rms} = 0.3$. The forecasts using
  the polarization estimator assume that one third of these galaxies are
  detected sufficiently well in polarization such that $\alpha_{\rm
    rms} = 7$ degs. The cosmic variance limit is also shown for
  comparison.}
\label{fig:cls}
\end{figure}

In terms of forecasting constraints for future surveys,
\cite{kaiser98} has shown that the errors obtainable on the shear
power spectrum with a survey covering a fraction of sky $f_{\rm sky}$
and with a galaxy number density $\bar{n}$ are given by
\be
\Delta C_\ell = \sqrt{\frac{2}{(2\ell + 1)f_{\rm sky}}} \left(C_\ell +
\frac{\vepsilon_{\rm rms}^2}{\bar{n}} \right).
\ee
The equivalent expression for the polarization case is
\be
\Delta C_\ell = \sqrt{\frac{2}{(2\ell + 1)f_{\rm sky}}} \left(C_\ell +
\frac{16 \alpha_{\rm rms}^2 \vepsilon_{\rm rms}^2}{\bar{n}_{\rm pol}} \right).
\label{eqn:cls_forecast_pol}
\ee
In Fig.~\ref{fig:cls}, we have plotted these predictions for some
representative numbers chosen to mimic large-scale surveys to be
conducted with the SKA. If the polarization properties of star-forming
galaxies assumed here turn out to be reasonable, then it will be very
interesting to apply the polarization techniques described here and
elsewhere to forthcoming high-resolution $\mu$Jy-sensitivity radio
surveys.

\bibliographystyle{apj}

\end{document}